\documentclass[twocolumn]{aastex62}

\usepackage{amsmath}
\usepackage{chngcntr}

\graphicspath{{./}{figures/}}

\received{December 2, 2019}
\revised{January 23, 2020}
\accepted{January 27, 2020}
\submitjournal{ApJL}

\shorttitle{A low-mass companion in MWC 297}
\shortauthors{Ubeira Gabellini M. Giulia}

\begin{document}

\title{Discovery of a low-mass companion embedded in the disk of the young massive star MWC~297 with VLT/SPHERE\footnote{Based on observations performed with ESO Telescopes at Paranal Observatory under programs 095.C-0787 and 0101.C-0350.}}

\correspondingauthor{Maria Giulia Ubeira Gabellini}
\email{maria.ubeira@unimi.it}

\author[0000-0002-0786-7307]{M. Giulia Ubeira-Gabellini}
\affil{Dipartimento di Fisica, Universit{\`a} Degli Studi di Milano, \\Via Celoria, 16, Milano, I-20133, Italy.
 \\
}
\affil{European Southern Observatory,\\ Karl-Schwarzschild-Str. 2,\\ D-85748 Garching bei M\"unchen, Germany.\\
}
\nocollaboration

\author{Valentin Christiaens}
\affiliation{School of Physics and Astronomy, \\Monash University, Clayton VIC 3800, Australia\\
}
\nocollaboration


\author{Giuseppe Lodato}
\affiliation{Dipartimento di Fisica, Universit{\`a} Degli Studi di Milano, \\Via Celoria, 16, Milano, I-20133, Italy.
 \\
}
\nocollaboration

\author{Mario van den Ancker}
\affiliation{European Southern Observatory,\\ Karl-Schwarzschild-Str. 2,\\ D-85748 Garching bei M\"unchen, Germany.\\
}
\nocollaboration

\author{Davide Fedele}
\affiliation{INAF - Osservatorio Astrofisico di Arcetri,\\
Largo E. Fermi 5, I-50125 Firenze, Italy.}
\nocollaboration

\author{Carlo F. Manara}
\affiliation{European Southern Observatory,\\ Karl-Schwarzschild-Str. 2,\\ D-85748 Garching bei M\"unchen, Germany.\\
}
\nocollaboration

\author{Daniel J. Price}
\affiliation{School of Physics and Astronomy, \\Monash University, Clayton VIC 3800, Australia\\
}
\nocollaboration



\begin{abstract}
We report the discovery of a low-mass stellar companion around the young Herbig Be star MWC 297. We performed multi-epoch high-contrast imaging in the near infrared (NIR) with the Very Large Telescope (VLT)/Spectro-Polarimetric High-contrast Exoplanet REsearch (SPHERE) instrument. The companion is found 
at projected separation of 244.7$\pm$13.2~au and a position angle of 176.4$\pm$0.1 deg. The large separation supports formation via gravitational instability. 
From the spectrum,
we estimate a mass of 0.1--0.5~M$_{\odot}$, the range conveying uncertainties in the extinction of the companion and in evolutionary models at young ages. 
The orbit coincides with a gap in the dust disk inferred from the Spectral Energy Distribution (SED).
The young age ($\lesssim$ 1 Myr) and mass ratio with the central star ($\sim 0.01$) makes the companion comparable to PDS 70~b, suggesting a relation between formation scenarios and disk dynamics. 
\end{abstract}



\section{Introduction} \label{sec:intro}

Binary star formation theories such as disk fragmentation \citep{Bonnell1994}, capture \citep{Tohline2002}, or core fragmentation \citep{Bonnell1991} are best tested with direct imaging of young objects. But the number of low-mass companions around pre-main-sequence stars detected by direct imaging remains low  \citep[e.g.][]{Bowler2016}.  
The situation is improving thanks to purpose-built high-contrast instruments such as the
Spectro-Polarimetric High-contrast Exoplanet REsearch instrument (SPHERE, \citealt[][]{Beuzit2008}) at the Very Large Telescope (VLT) and Gemini Planet Imager (GPI, \citealt[][]{Macintosh2014}). Using these new instruments, 
\citet[][]{Keppler2018} detected and confirmed a companion within the gap of the transition disk around PDS~70. 
\\

 In this Letter we report the discovery of a low-mass companion in the disk around Herbig Be star MWC~297 using high-contrast observations with VLT/SPHERE-IFS. 

\section{MWC 297} \label{MWC297}
MWC 297 (RA(J2000)~=~18~27~39.527, Dec(J2000)~=~-03~49~52.05) is a young pre-main-sequence ($<1$Myr) Herbig Be star (spectral type B1.5), with M$_{\star}\sim$17~M$_\odot$ \citep{Vioque2018} located in the L515 region at a distance of $\sim$ 375 pc (\citealt{Vioque2018}, Gaia DR2). It was classified as a Class II, Group I source \citep{Meeus2001} from SED fitting \citep{Mannings1994}. 
From the mm-spectral slope of the SED, \citet{Manoj2007} argued for either a compact disk or for grain growth in the circumstellar environment.
The system has a compact circumstellar disk (\citealt{Weigelt2011}, Br$\gamma$ and NIR continuum visibilities study) 
and with low inclination ($\sim$5$^{\circ}$, \citealt{Alonso2009}).
Finally, \citet{Alonso2009} observed the disk at millimeter wavelengths with the Very
Large Array. They modeled the SED using a two-component disk, with inner ($\sim$7.5 to 43.5 au) and outer ($\sim$300 to 450 au) parts and a gap in between. Both the presence of a companion or grain growth in the outer disk may explain such a gap. 
The authors ruled out a companion due to the apparent non-detection of any point-like source at the suggested distance ($\sim$270~au when rescaled to Gaia distance).  


\begin{table}
	\caption{Physical properties of MWC 297}
	\label{tab:litsource}
   	\addtolength{\tabcolsep}{-2.pt} 
    \begin{tabular}{p{30cm}p{30cm}} 
	\begin{tabular}{cccc} 
	\hline
	\hline
\multicolumn{1}{c}{Param.} & Units & Value & References\\
\hline
 \multicolumn{1}{c}{$d$} & pc & 375$\pm$20 & \citet[][]{Gaia2018} \\
\multicolumn{1}{c}{Age} & Myr & $<1$&  \citet[][]{Acke2006}\\
\multicolumn{1}{c}{} &  & &  \citet[][]{Vioque2018}\\
\multicolumn{1}{c}{Sp.T.} &&B1.5Ve & \citet[][]{Drew1997}\\
\multicolumn{1}{c}{Group} &&I & \citet[][]{Meeus2001}\\
\multicolumn{1}{c}{$T_{\rm eff}$} & K & 23700 & this work\\
\multicolumn{1}{c}{$A_{\rm v}$} & mag & 7.72 & this work\\
\multicolumn{1}{c}{$\log(L_{\rm bol}$)} &L$_{\odot}$ & 4.59 &\citet[][]{Vioque2018}\\ 
\multicolumn{1}{c}{$M_*$} & M$_\odot$ & 16.9 & \citet[][]{Vioque2018}\\
\multicolumn{1}{c}{$R_*$} & R$_\odot$ & 9.17 & this work\\
	\hline
	\hline
	\end{tabular}
	\end{tabular}
\\
{\bf Notes. }{\small $d$: Gaia distance; Sp.T.: spectral type; Group: Disk classification according to \citet{Meeus2001}; $T_{\rm eff}$: effective temperature, $A_V$: extinction, $L_{\rm bol}$: bolometric luminosity; $M_*$: stellar mass; and $R_*$: stellar radius.
}
\end{table}

\section{Observations and data reduction}\label{obs}
\subsection{Derivation of stellar properties}
Table~\ref{tab:litsource} summarizes the stellar and disk properties. The effective temperature and interstellar extinction were derived following \citet{vandenAncker1998}: the observed SED (between 0.3 - 1.2 $\mu$m) was fitted using atmospheric models of \citet{Kurucz1991} and the dereddening law from \citet{Cardelli1989} with R$_{\rm v}$=3.1. 
The stellar radius was estimated based on $L_{\rm bol}$ and T$_{\rm eff}$.

\subsection{Observations}

We observed MWC~297 on April 29, 2015 and on June 28, 2018 with SPHERE 
in the IRDIFS-EXT mode i.e. simultaneous Integral Field Spectroscopy (IFS) in the $YJH$ bands, and dual-band imaging in the $K$ band. The first observation (2015-04-29) was taken in field tracking mode, while for the second set (2018-07-28) we used the pupil tracking mode (Table \ref{tab:compprop}). 
The IFS data are cubes of 39 monochromatic images in the NIR encompassing a Field of view of 1.$^{\prime \prime}$73 x 1.$^{\prime \prime}$73. The spectral resolution was R$\sim$30 for the IRDIFS-EXT mode ($Y$-$H$, 0.95$<\lambda<$1.65$~\mu$m). 
The N-ALC-YJH-S coronagraph (inner working angle $\sim$0.$^{\prime \prime}$15) was used.

We obtained ``Flux" and ``Star Center" calibration images at the beginning and end of both observing sequences. 
The ``Flux" images 
were obtained by offsetting the central star from the coronagraphic spot and used to measure the unsaturated peak flux of the star. 
The ``Star Center" images allowed us to measure the position of the star behind the coronagraph, located at the center of the four replicas produced by the adaptive optics system. 

\subsection{Data reduction}\label{datared}
We used the ESO pipeline\footnote{http://www.eso.org/sci/software/pipelines/sphere/; v0.24.0, for the first dataset; v0.36.0, for the recent data.} 
to reduce the IFS data. 
We used a function implemented in the Vortex Imaging
Pipeline\footnote{https://github.com/vortex-exoplanet/VIP.} (VIP, \citealt{Gomez-Gonzalez2017}) to correct
for clumps of bad pixels through an iterative sigma filtering process. 
For the centering, we increase the signal-to-noise (SNR) of the star replicas using a high-pass filter that subtracts the image itself with a median low-pass filtered version of the image.
We fitted the four replicas with a 2D Moffat function (in VIP) to derive the centroid of the star in each frame. 
We then interpolated the values of the derived center (taken at the beginning and end of the observations) considering the observation time of the science images. 
Finally, the error was considered to be the discrepancy in the value between two sequential sets of center images (on average $\sigma_{x}$=0.04 pixels and $\sigma_{y}$=0.08 pixels).

\begin{figure*}[htb!]
\centering
\includegraphics[width=\hsize, angle=0]
{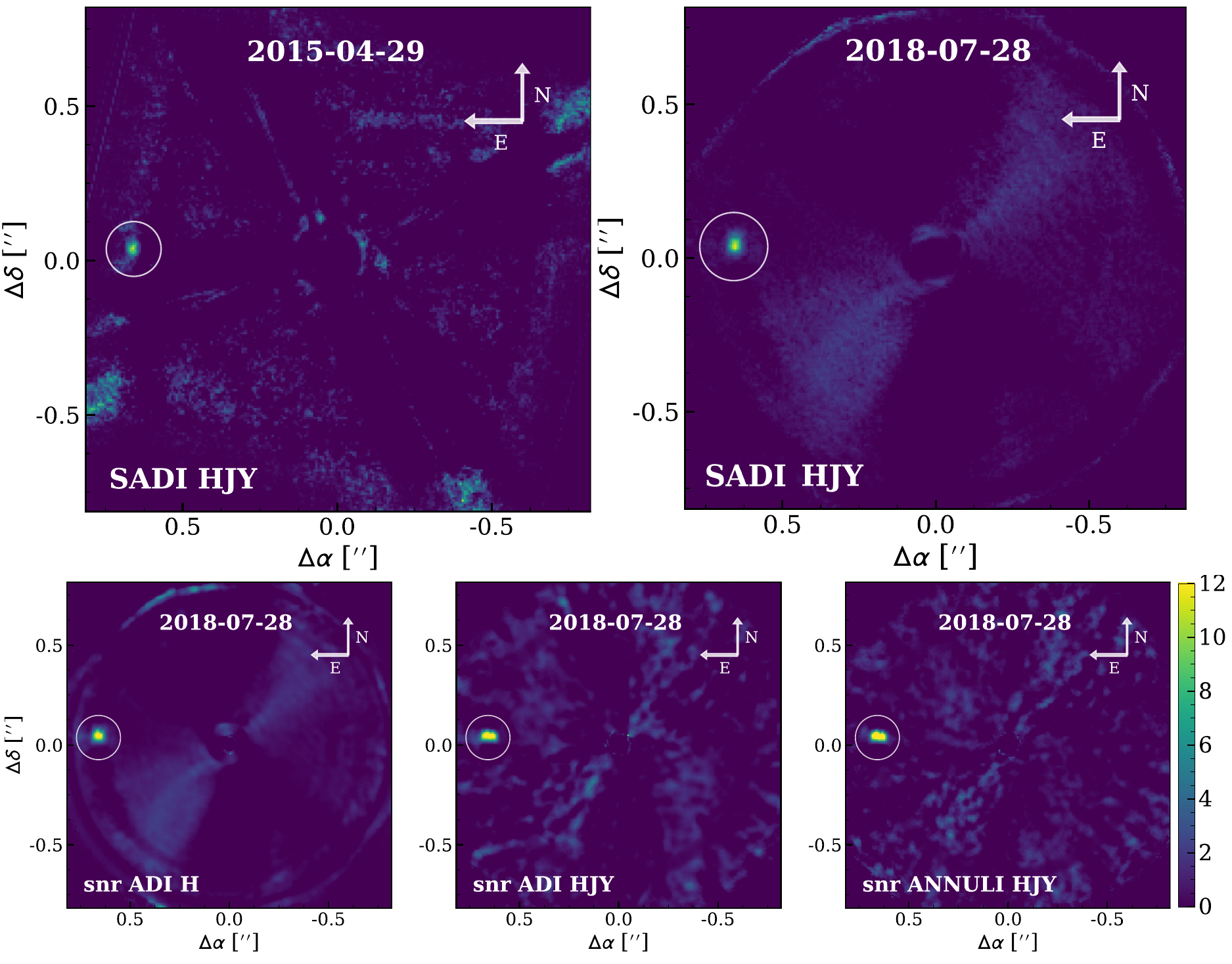}
\caption{Top: MWC~297B detected with SADI in 2015 (left) and 2018 (right) combining all wavelengths. 
Bottom: SNR (day: 2018-07-28) using ADI performed in full frame with just H-Band frames (left), with all frames combined (center) and done in annuli (right).
The companion is always detected (white circle), irrespective of technique. We also detected the companion in J and Y Band images with ADI, but fainter. 
}
\label{fig:mwc297_images}
\end{figure*}

\subsection{Post-processing using VIP}
Calibrated frames are still affected by quasi-static speckles produced by the star \citep{Marois2006}. 
Speckles move radially with wavelength, while real features remain fixed (spectral information). This is key to the Spectral Differential Imaging (SDI) algorithm (e.g., \citealt[][]{Sparks2002}). Also, fixing the pupil of an altitude-azimuth telescope during an observing sequence, most quasi-static speckles remain fixed in the image, while real features rotate (angular information). Angular Differential Imaging (ADI; e.g., \citealt[][]{Marois2006}) is based on this idea.
The IFS cubes contain the spectral information, while angular information is available when the rotator is moved to maintain the pupil fixed. 
We used principal component analysis (PCA)-based algorithms in VIP to model and subtract the stellar point-spread function (PSF) and associated speckles. 
For both sets of observations we applied PCA-SDI, using the spectral information alone, and PCA-SADI, 
where the PCA library was built using both the angular and spectral information (\citealt[][]{Pueyo2012}). We also tested the algorithm in two separate steps \citep[PCA-SDI~+~PCA-ADI;][]{Christiaens2019b}, but obtained noisy final images.
For the second observational set, we also used another algorithm: PCA-ADI, using only the angular information, performed either in full frames \citep{Soummer2012} or in concentric 2-FWHM wide annuli on individual spectral channels \citep{Absil2013}. 
\section{Characterization of the companion}\label{planetmass}

\subsection{2015 detection}
We detected a bright companion in the outer disk of MWC 297 on 29th~April~2015 located $\sim$246.4~au from the central star. The detection was obtained using the PCA-SADI (Fig.~\ref{fig:mwc297_images}, top left) and PCA-SDI techniques, with 4$\sigma$ and 5$\sigma$ significance. 
The companion was detected in the averaged $H$-band image (SNR$\gtrsim$4), but not in $J$ and $Y$. 

\subsection{2018 detection}
We performed follow-up observations with longer integration time (Table \ref{tab:compprop}) on 28th July 2018. We re-detected the companion in $H$ and $J$-bands with four different post-processing methods 
(SNR$>$4). We detected it also in the $Y$-band just using ADI. 
Figure~\ref{fig:mwc297_images} shows that the point-like source is detected regardless of the post-processing method (SADI, ADI, ANNULI and SDI --- not shown here) and of wavelength ($H$, $J$ and $Y$ bands all show the companion). 

\subsection{Spectro-astrometry}\label{spec-astr}
We used the Markov Chain Monte Carlo (MCMC), a nested sampling algorithm coupled to the negative fake companion technique implemented in VIP to derive the position and the flux of the companion at each wavelength (e.g.~\citealt[][]{Marois2010,Wertz2017}). We first estimated the position and flux using the Nelder-Mead simplex-based algorithm \citep[][]{Nelder1965}, and then fed these first estimates to the MCMC routine. A negative PSF was injected in the original data cube in order to completely delete the signal of the real companion measured in the final PCA-ADI post-processed image. The process produces a posterior distribution of the three parameters and stops upon convergence to minimal absolute residuals in an aperture centered on the location of the companion.
Finally, this routine gave us the companion separation, position angle and flux, with errors (Table~\ref{tab:compprop}).

\begin{figure*}[htb!]
\centering
\includegraphics[width=\hsize, angle=0]{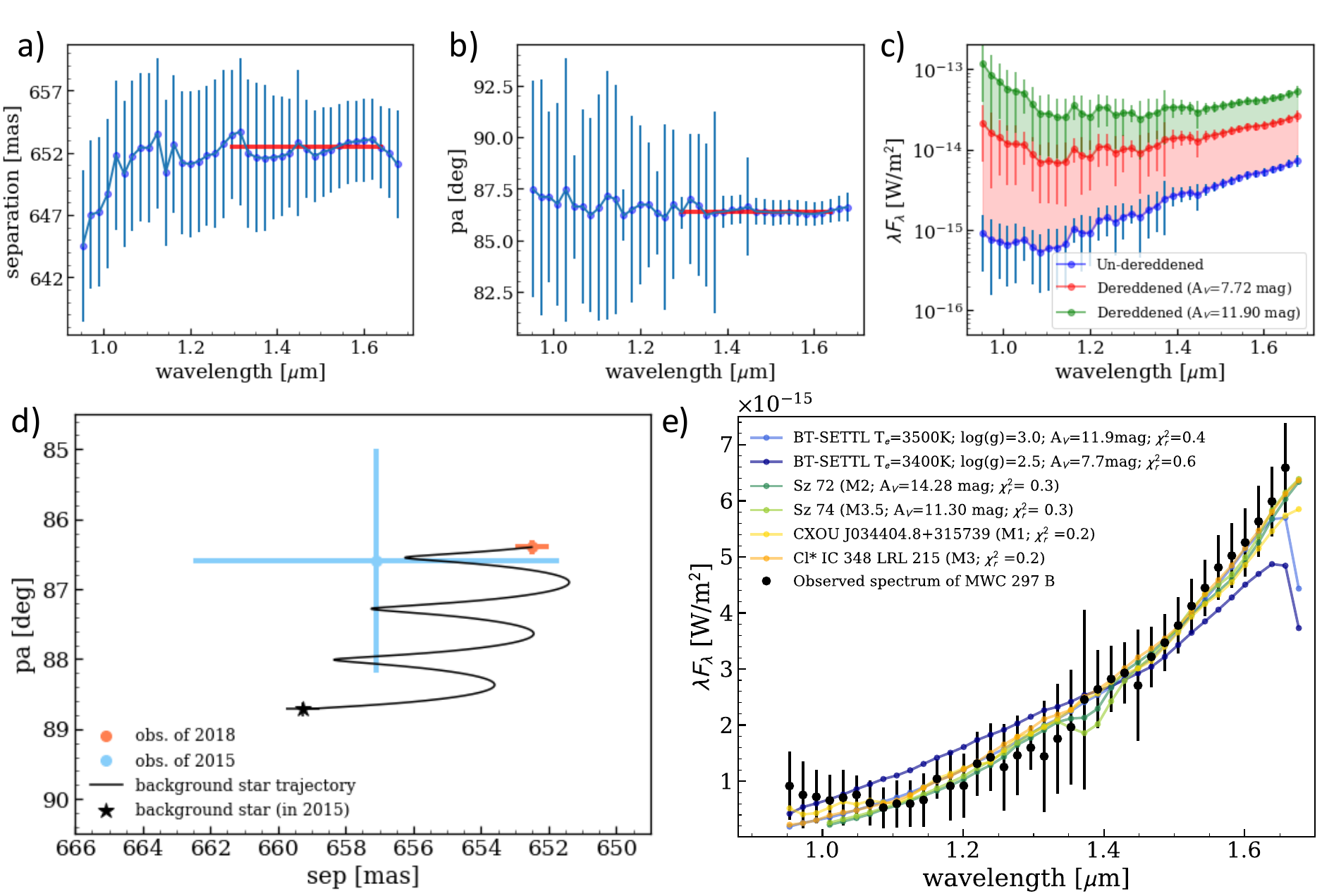}
  \caption{Top: MCMC fit (on second epoch) for separation (left) and position angle (center). Red lines show weighted average, inferred where the SNR is highest (1.29$\mu$m$\lesssim\lambda\lesssim$1.64$\mu$m). 
Top right: IFS spectrum of companion in physical units, shown undereddened (\emph{blue points}), dereddened with the stellar extinction (\emph{red points}), and with $A_{V}$=11.9~mag (\emph{green points}). Bottom left: Companion astrometry on the two datasets with their 1$\sigma$ uncertainties (first epoch: light blue, second epoch: orange; Table \ref{tab:compprop}). 
Black line shows the trajectory for a background star going back to the 2015 epoch. 
 Bottom right: Best-fit BT-SETTL models with extinction as free parameter (light blue) and with $A_V$ inferred from the central star (dark blue); best-fit YSOs (green) and SpeX template spectra (yellow).
 } 
  \label{fig:mwc297_combined}
\end{figure*}


\subsubsection{Astrometry} \label{Astrometry1}
Using MCMC on the second epoch, we derived the position ($r$ and PA) and relative error of the companion 
for each wavelength and computed the weighted average (Figure~\ref{fig:mwc297_combined}, panel a, b; red line). For the first epoch, it was not possible to use MCMC. We therefore fitted a 2D Gaussian to derive the position and took the weighted average of the results of the SDI and SADI methods.
We considered a pixel scale of $7.46\pm0.02$ mas/pixel \citep{Maire2016}. 
The separation uncertainty was computed as a sum in quadrature of the uncertainties on the stellar and companion position and on the pixel scale for each frame (Figure~\ref{fig:mwc297_combined}, panel a; Table \ref{tab:compprop}). We took into account the target distance error to derive the separation in au (Table \ref{tab:compprop}).

The position angle (PA) is affected by the error on the True North angle determination of -102.18$\pm$0.13 deg \citep{Maire2016}, used to derive the astrometry. We propagated the errors in the position and true north to get the final error (panel b, Figure~\ref{fig:mwc297_combined}, Table \ref{tab:compprop}). 

Figure~\ref{fig:mwc297_combined} (panel d) suggests a companion comoving with the host star on a trajectory more consistent with Keplerian motion
. A background star would move on the trajectory shown by the black line: its proper motion after 3.25 yr exceeds the centroid discrepancy of the two observation sets (orange and blue), albeit within 2$\sigma$ uncertainty. Assuming a face-on circular orbit (recalling the $\sim$5$^{\circ}$ disk inclination), Keplerian motion would account for a shift in position angle of 1.26$^{\circ}$
, inside the error bar of the first detection.


Our two-epoch astrometry alone does not rule out the possibility of a background object. Therefore, we used the TRILEGAL model of the Galaxy to estimate the probability of being a background star \citep{Girardi2005}. TRILEGAL yields 6553 stars with an $H$-band apparent magnitude brighter or equal to that of our companion candidate ($H~\leq~13.84$~mag; Sec.~\ref{SpectroPhotom}) within a 30\arcmin~$\times$~30\arcmin~patch of sky centered on the star, hence a density of 0.002 arcsec$^{-2}$. The probability is thus $1-\mathcal{P}(n=0|\lambda=0.002, B=4) \approx 0.8$\%, where $\mathcal{P}(\lambda,B)$ is the spatial homogeneous Poisson point process probability with rate $\lambda$ and area $B$. Given the separation of $\sim$0.\arcsec7, we conservatively consider a $2\arcsec\times2\arcsec$ box centered on the star for the area.

\subsubsection{Spectro-photometry}\label{SpectroPhotom}
The longer integration time of the 2018 data allowed us to detect the companion candidate at a significant level in all $Y$, $J$ and $H$ bands (Fig. \ref{fig:mwc297_images}, center left, shows H band), and to derive its spectrum (Figure~\ref{fig:mwc297_combined}, panel c).
For each spectral frame, we measured the flux from the star using the ``Flux" image and the companion flux using the MCMC method described in Section \ref{spec-astr}. The stellar flux error was considered as the discrepancy between two sets of ``Flux" images. 

To produce the final calibrated spectrum (Figure~\ref{fig:mwc297_combined}, panel c), we multiplied the measured spectrum of the companion by the ratio between the stellar flux in physical units, obtained though a polynomial fit of the stellar SED in the IFS wavelength range, and in ADUs in each spectral channel.
For completeness, we also measured the total emission of the companion over the star in bands $H$, $J$ and $Y$ with errors (Table \ref{tab:compprop}) and derived the apparent magnitude of the companion in those bands (13.86~mag, 16.21~mag and 17.30~mag, respectively).


\subsection{Spectral analysis} \label{SpectralAnalysis}

The undereddened spectrum of the companion (Fig.~\ref{fig:mwc297_combined}, panel c; \emph{blue points}) shows a very red slope, suggesting significant extinction on the companion, not necessarily the same of the star. 
Each component might be embedded and surrounded by their own disk, in addition to any remnant envelope \citep[e.g.][]{Bowler2014,Mesa2019}. 
Therefore, following \citet{Christiaens2018}, we considered extinction as a free parameter when fitting BT-SETTL models \citep{Allard2012}.
Our grid of BT-SETTL models contains four free parameters: Effective temperature, $T_{\rm eff}~\in~[1200K,5500K]$ in 100K steps; surface gravity, $\log(g)\in[2.5,5.0]$ in 0.5dex steps; radius, $R_B \in~[0.1~$R$_{\odot},3.5$R$_{\odot}]$ in 0.01 R$_{\odot}$ steps; and extinction, $A_V~\in~[0,21]$ mag in 0.1 mag steps. 
We then considered the same grid, but we fixed the extinction to 
$A_V$=7.72~mag (Figure~\ref{fig:mwc297_combined}, panel c, \emph{red} points), same as for the central star (Section \ref{MWC297}). 


Next, we considered two libraries of young stellar objects (YSOs) template spectra: (i) all 76 pre-main sequence stars spectra compiled in \citet{Alcala2014} and \citet{Manara2013,Manara2017}, which are members of the TW Hya, $\sigma$ Ori, Lupus I, III and IV star forming regions, spanning G5 to M8.5 spectral types; and (ii) all young dwarfs from the SpeX library \citep{Burgasser2014}, identified based on their gravity class or their membership to young ($<$ 10 Myr old) clusters. In either cases, we considered two free parameters to account for different A$_{\rm V}$ and distance between observed and template spectra. 


For all spectral fits, we convolved the models and templates with the IFS spectral response before binning them to the same wavelength sampling. We then minimized a goodness-of-fit indicator $\chi^2$ that accounts for the spectral covariance of the IFS instrument \citep{Greco2016,Delorme2017}.
 


Figure~\ref{fig:mwc297_combined} (panel e) shows the best-fit BT-SETTL and YSOs template spectra (\emph{blue} and \emph{green} color) with the undereddened spectrum of the companion candidate (\emph{black points}). With extinction as a free parameter, the best-fit BT-SETTL model has $T_{\rm eff} = 3500$ K, $\log(g) = 3.0$, $R_B = 1.13 R_{\odot}$ and $A_V = 11.9$ mag (\emph{solid line}; $\chi_r^2 \sim 0.4$), consistent with a young (very-low gravity), gravitationally contracting and embedded stellar mass companion surrounded by a lot of dust. 
By contrast, lower values of extinction (e.g. $A_V=$7.72 mag, \emph{dotted line}), gave significantly worse fits.

The best-fit template spectra correspond to early M-type (M1 to M3.5) YSOs from (i) the 1--3 Myr-old Lupus I cloud \citep[Sz 72 and Sz 74;][]{Alcala2014}; and (ii) the $\sim$2 Myr-old cluster IC~348 \citep[CXOU J034404.8+315739 and Cl* IC 348 LRL 215;][]{Luhman2003}. Interestingly, both the SpeX targets are 
located in the youngest part of the IC~348 cluster, where class 0/I objects have been identified \citep{Luhman2003,Luhman2016}. In particular, they could also be class 0/I objects given their significantly lower differential extinction compared to the best-fit extinctions associated with the Lupus I and BT-SETTL spectra, suggesting $A_V \gtrsim 10$ mag for the companion. 

Based on the empirical relationship between spectral type and effective temperature inferred in \citet{Luhman2003} for IC~348, spectral types M1--M3.5 would correspond to $T_{\rm eff}$ = 3350--3700~K, consistent with our best-fit BT-SETTL model effective temperature.

\begin{table*}[t!]
	\caption{Observation log and MWC~297~B properties. 
	}
	\label{tab:compprop}
 	\addtolength{\tabcolsep}{-1.pt} 
 	\begin{tabular}{cccccccccccc}
	\hline
	\hline
	 \multicolumn{1}{c}{Obs.~date$^{\dagger}$} & Exp. & Track. & seeing & $\Delta$ PA  & sep  & sep &PA & $\Delta H$ & $\Delta J$ & $\Delta Y$ & $M_B$\\ 
	\multicolumn{1}{c}{}& [s] & & & [deg] & [mas]  & [au] & [deg] & [mag] & [mag] & [mag] & [M$_{\odot}$]\\ 
	\hline
	\multicolumn{1}{c}{2015-04-29}& 1664 &F &0.68 & 1.1 & 657.1$\pm$5.4& 246.4$\pm$15.2 &176.6$\pm$1.6 &  10.19$\pm$0.53& -- & -- & 0.1--0.2\\
	\multicolumn{1}{c}{2018-07-28} &5760 & P & 0.91 & 54.3 & 652.5$\pm$0.5 & 244.7$\pm$13.2 & 176.4$\pm$0.1 & 9.49$\pm$0.03 & 10.37$\pm$0.3 & 10.23$\pm$0.10 & 0.1--0.50 \\
	\hline
 	\end{tabular}
	\\
	{\bf Notes. } {\small $^{\dagger}$Programs 095.C-0787 (PI: van den Ancker) and 0101.C-0350 (PI: Ubeira Gabellini), respectively. Table lists observation date, total integration time, telescope tracking mode (F: field-tracking; P: pupil-tracking), mean seeing, total field rotation, separation (mas and au), position angle, delta magnitude (H, J, Y) and estimated companion mass. 
	}
\end{table*}
\subsection{Mass estimate}
Considering an extinction of $A_V$ = 11.9 $\pm 1.0$ mag (Sec.~\ref{SpectralAnalysis}), our de-reddened $J$- and $H$-band absolute magnitudes are 5.3$\pm 0.3$ mag and 4.2$\pm 0.1$ mag (using \citealt[][]{Cardelli1989}), respectively.
We compared the absolute magnitudes and colors with BCAH98, AMES-Cond and BT-SETTL (\citealt[][]{Chabrier2000,Baraffe1998,2003IAUS..211..325A}) models that suggests a mass of $\sim$0.10--0.25~M$_{\odot}$. Comparing the T$_{\rm eff}$ and age with stellar isochrones \citep{Baraffe2015} suggest a mass of 0.25--0.5~M$_{\odot}$ (Table~\ref{tab:compprop}). 
Considering that this estimate assumes an age of 1~Myr --- the youngest available, but an upper limit for MWC~297 (\citealt{Vioque2018} suggest $\approx$0.02--0.03~Myr) --- the companion mass may be lower. 
This is consistent with the best fit YSOs and SpeX template spectra with mass $\sim$ 0.45-0.50 M$_{\odot}$, targets older (1--3~Myr) than MWC 297. 
For the 2015 epoch, we estimated the mass using only the dereddened absolute $H$-band magnitude, due to the lack of obvious detection in other bands.

\section{Discussion}\label{discussion}

The $0.8$\% probability of being a background star (Sec.~\ref{Astrometry1}) suggests that the detected point source is a bound companion to MWC~297. Our spectral analysis (Sec.~\ref{SpectralAnalysis}) further argues in favor of a young and embedded early M-dwarf.
BT-SETTL models are uncertain at low gravity \citep[e.g.][]{Bonnefoy2014}, and the template library lacks spectra younger than $1~$Myr old, both suggesting an even less massive object. Moreover, the spectral fit is not able to reproduce exactly the observed spectrum (Fig.~\ref{fig:mwc297_combined}, panel e). Using dust extinction curves different from those assumed for the ISM may also improve the fit \citep[e.g.][]{Marocco2014}.
Furthermore, the very red slope possibly is partially due to excess dust thermal emission from a circum-secondary disk \citep[e.g.][]{Christiaens2018} - with less extinction needed. Follow-up observations at longer wavelengths are required to better refine the characteristics of the companion and test the presence of a hot circum-secondary disk component.

Our detected low-mass companion might be carving the gap in dust thermal emission suggested by \citet{Alonso2009}, based on the SED and 1.3~mm and 2.6~mm IRAM Plateau de Bure (PdBI) interferometer data. The resolution of the PdBI data (1.$^{\prime\prime}$1 $\times$ 0.$^{\prime\prime}$4 for 1.3~mm and 1.$^{\prime\prime}$4 $\times$ 0.$^{\prime\prime}$9 for 2.6~mm), however, was too coarse to resolve the 0.$^{\prime\prime}$65 separation between the central star and the source. Atacama Large Millimeter/submillimeter Array (ALMA) submillimeter continuum observations would allow to test whether the companion lies within a large annular gap.

\begin{figure}[htb!]
\centering
\includegraphics[width=\hsize, angle=0]{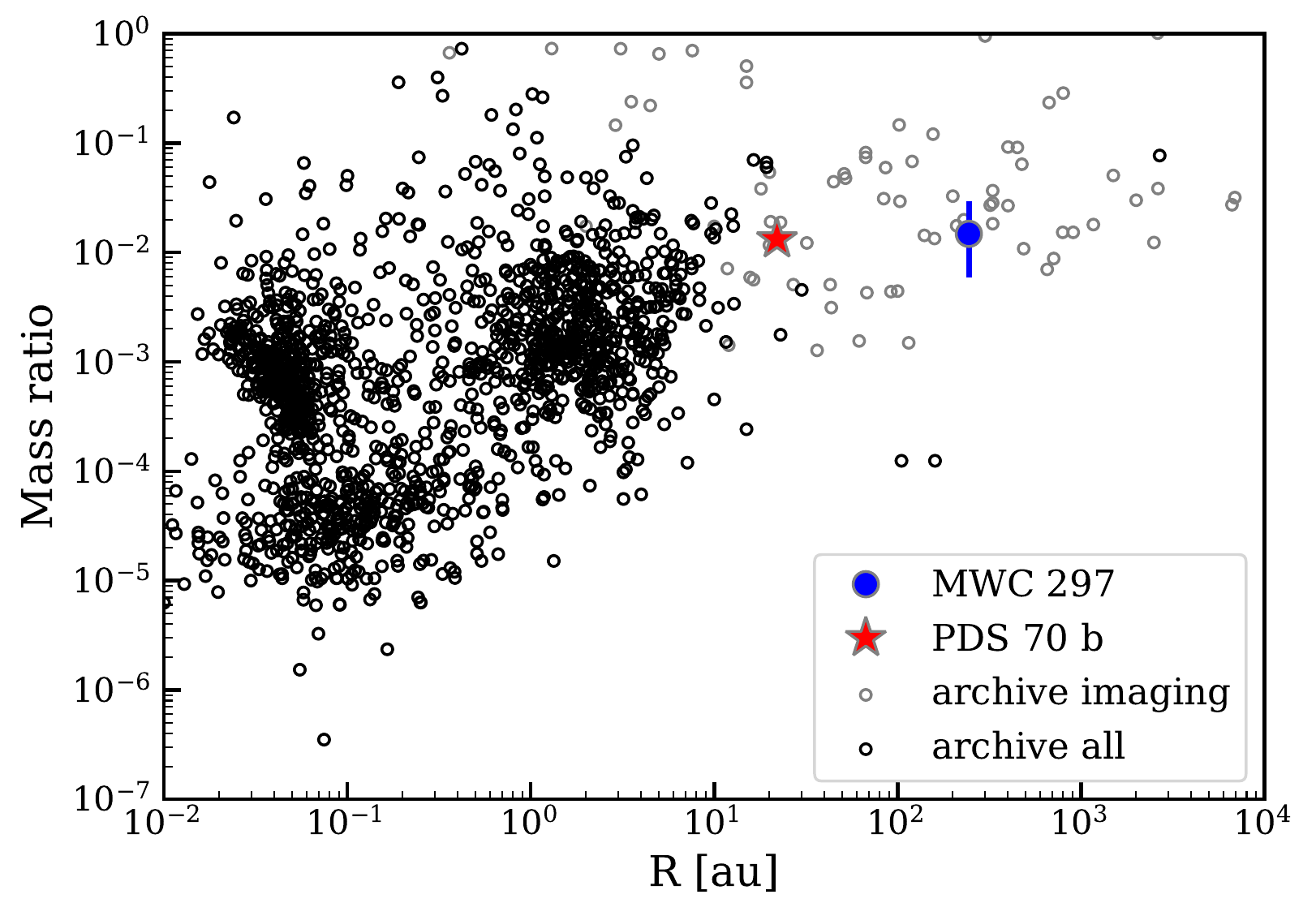}
\includegraphics[width=0.97\hsize, angle=0]{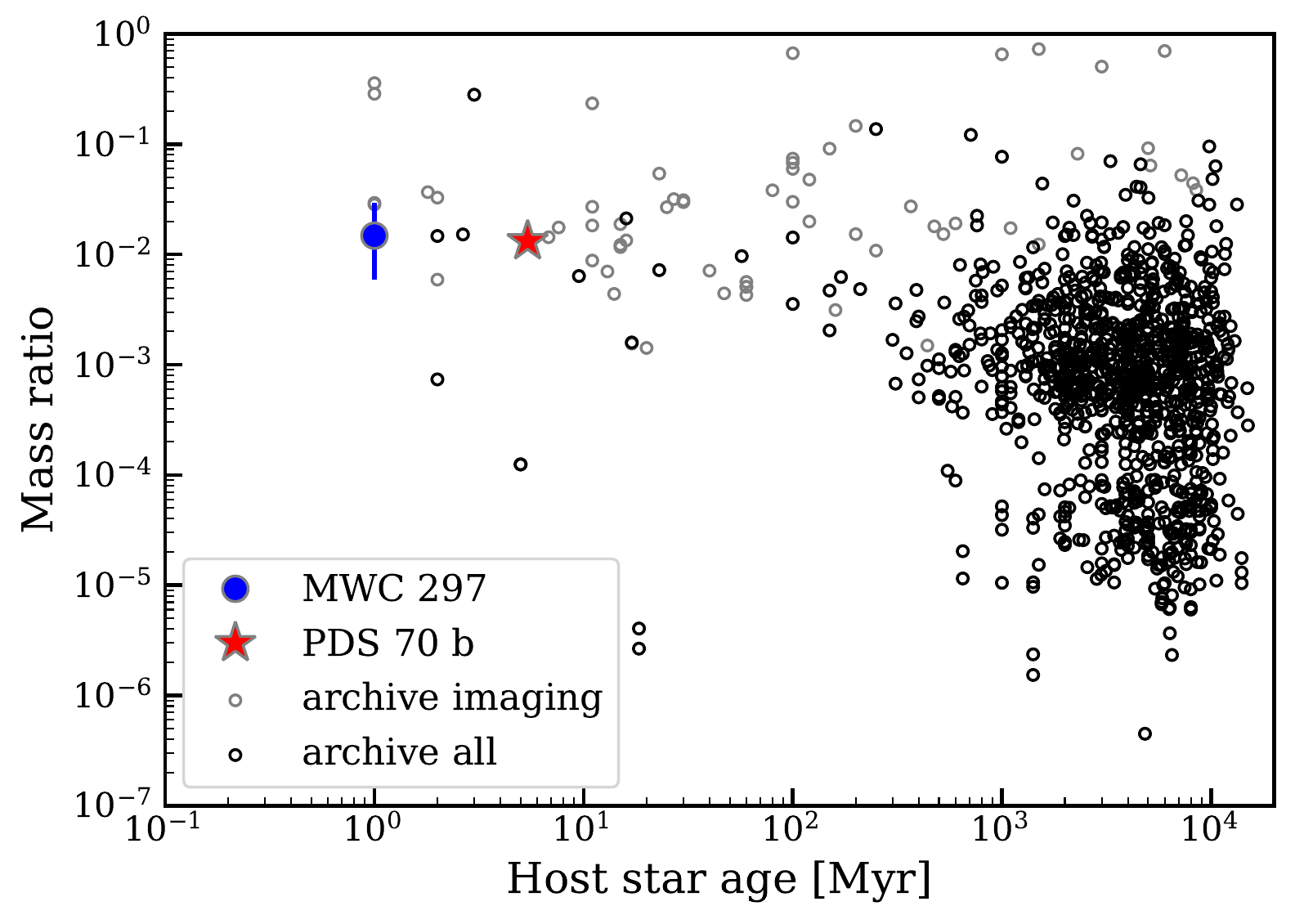}
\caption{Ratio between companion and host stellar mass (y-axis) vs orbital separation (top) or age of stellar host (bottom). 
Empty circles show known exoplanets from direct imaging (gray); and other methods (black). Blue circle shows our companion MWC 297~B; red star shows PDS 70~b  \citep{Keppler2018}. A companion mass of 0.25~$\substack{+0.25 \\ -0.15}$~M$_{\odot}$ implies a mass ratio similar to that of PDS 70~b. 
}
\label{sep_vs_mratio}
\end{figure}
Figure~\ref{sep_vs_mratio} compares our companion detection to archival data from the exoplanets.eu database assuming a companion mass of 0.25~$\substack{+0.25 \\ -0.15}$~M$_{\odot}$. Our target is low-mass compared to the host star and at large separation, similar to other direct imaging detections. The $\sim$10$^{-2}$ mass ratio is similar to that of PDS 70~b.
Interestingly, the companion around MWC 297 is one of the few discovered around young host stars (bottom panel). Most archival companions with ages below 10 Myr found with direct imaging are yet to be confirmed. 


Our best-fit extinction is high, but similarly embedded young low-mass companions have been detected e.g. FW~Tau~C \citep{Bowler2014} and R~CrA~B \citep{Mesa2019}. 
It may have an edge-on disk (like TWA~30~B and FW~Tau~C; \citealt{Looper2010,Wu2017}). Follow-up with ALMA is required to confirm this for MWC~297~B. 


\section{Summary and conclusions}\label{conclusion}
We detected MWC~297~B in $H$ band on 2015-04-29 and again in $Y$, $J$ and $H$ bands on 2018-07-28. Astrometry favors a gravitationally bound object. Spectral characterization suggests a young ($<$1~Myr) low-mass companion
(0.25~$\substack{+0.25 \\ -0.15}$~M$_{\odot}$) and high extinction ($A_V\sim$11.9~mag). The large separation supports formation via gravitational instability. 
The mass ratio is comparable to that of PDS 70~b, but in the stellar mass regime, suggesting a similar formation process for low-mass companions around high- and low- mass stars. Finally, the companion could be responsible for the dust gap inferred by \citet{Alonso2009}.

\section*{Acknowledgements}
 We thank the anonymous referee for providing insightful comments. We used the SPHERE Data Centre, operated by OSUG/IPAG (Grenoble), PYTHEAS/LAM/CeSAM (Marseille), OCA/Lagrange (Nice) and Observatoire de Paris/LESIA (Paris) with funding from Labex OSUG@2020 (Investissements d’avenir ANR10 LABX56). We thank P. Delorme, S. Moehler, M. Reggiani and E. Sissa for useful discussions. DF acknowledges funding from the Italian Ministry of Education, Universities and Research, project SIR (RBSI14ZRHR). GL, MGUG and CFM received funding from the European Union Marie Skodowska-Curie grant 823823 (RISE DUSTBUSTERS project). VC and DP acknowledge Australian Research Council funding via DP180104235. CFM acknowledges an ESO fellowship.

\bibliography{mwc297_rev_comm_publ.bib}




\end{document}